\documentclass{emulateapj}
\usepackage{apjfonts}
\usepackage{epsfig}
\usepackage{graphicx}
\usepackage[usenames,dvips]{color}

\def\aap{ A\&A}

\def\apjs{ApJ Supp}
\def\apjl{ ApJL}

\begin{document}
\title{ Formation of black-hole X-ray binaries in globular clusters}
\author{N.\ Ivanova$^1$, S.\ Chaichenets$^2$, J.\ Fregeau$^3$, C.\ O.\ Heinke$^1$, J.C.\ Lombardi Jr.$^4$, T.\ Woods$^1$}
\altaffiltext{1}{University of Alberta, Dept. of Physics, 11322-89 Ave, Edmonton, AB, T6G 2E7, Canada}
\altaffiltext{2}{University of Alberta, Dept. of Mathematical and Statistical Sciences, CAB, Edmonton, AB, T6G 2G1, Canada}
\altaffiltext{3}{Chandra/Einstein Fellow; Kavli Institute for Theoretical Physics, UCSB, Santa Barbara, CA 93106, USA}
\altaffiltext{4}{Allegheny College, Meadville, PA 16335, USA}

\begin{abstract}
Inspired by the recent 
identification of the first candidate
BH-WD X-ray
binaries, where the compact accretors {\it may be} stellar-mass black hole candidates in extragalactic globular clusters, 
we explore how such binaries could be formed in a
dynamical environment.
We provide analyses of the formation rates via well known 
formation channels like binary exchange and physical collisions
and propose that the only possibility to form BH-WD binaries is
via coupling these usual formation channels with subsequent hardening
and/or triple formation.
Indeed, we find that the most important mechanism to make a BH-WD
X-ray binary from an initially dynamically formed BH-WD binary is
triple induced mass transfer via the Kozai mechanism.
Even using the most optimistic estimates for the formation rates,
we cannot match the observationally inferred production rates if black holes undergo significant 
evaporation from the cluster or form a completely detached subcluster of black holes.
We estimate that at least 1\% of all formed black holes, or presumably 10\% of the black holes present in the core now, must
be involved in interactions with the rest of the core stellar population.
\end{abstract}

\keywords{
X-rays: binaries  --- star clusters: general --- stars: kinematics and dynamics.
}

\section{Introduction}
\label{intro} 

The question of whether black holes (BHs) are present in globular clusters (GCs) has been 
discussed extensively  in the literature \citep[e.g.,][]{1993Natur.364..421K, 1993Natur.364..423S, 2000ApJ...528L..17P,2002MNRAS.330..232C}. 
The most obvious way to detect a BH in a GC is if the BH is in an X-ray binary.
\cite{kalogera_04} showed that, if a BH X-ray binary is formed dynamically 
by an exchange interaction in the core of a dense globular cluster,  it is very unlikely to be detected,   
as the duty cycle for such binaries is extremely low.
Tidally captured BH binaries, in contrast, would be {\it continuously luminous}, and the lack of 
{\it Galactic BH X-ray binaries} 
implies that {\it any such tidal captures destroy the} potential companions. 
We note that only non-degenerate donors were considered by \citet{kalogera_04}.
Indeed, in Galactic GCs, where the total number of LMXBs is relatively
small, no BH X-ray binary has been found so far \citep[e.g.,][]{2006csxs.book..341V}).
Among low-mass X-ray binaries (LMXBs) in the globular clusters of early-type galaxies, 
however, likely BH X-ray binary candidates have been reported 
\citep[e.g.,][]{2001ApJ...557L..35A,2002ApJ...570..618D,2002ApJ...574L...5K,2003ApJ...595..743S, kim06}. 
 The strongest such candidates have 
X-ray luminosities $L_{\rm X} >10^{39} \ {\rm ergs\ s^{-1}}$, well above the Eddington limit 
for a neutron star accreting helium, and are known as ultraluminous X-ray sources (ULXs).

A particularly interesting ULX has been identified in a globular cluster in NGC 4472.
It has $L_{\rm X}\sim 4 \times 10^{39} {\ \rm ergs \ s^{-1}}$ and has shown strong variability 
(likely due to variable absorption), indicating that it is a single source and thus likely a black hole X-ray binary \citep{2007Natur.445..183M}.
Keck spectroscopy of the globular cluster (RZ 2109) associated with this source
identified strong, broad (2000 km/s) [O III] emission lines, interpreted as an 
outflow from a stellar mass black hole accreting above or near its Eddington limit
 \citep{2008ApJ...683L.139Z}.  
The low H$\alpha$/[O III] ratio suggests a hydrogen-poor (white dwarf) donor, 
while the assumption that the breadth of the line is due to a wind driven 
from near-Eddington accretion implies a BH  mass of $5-20 M_\odot$ \citep{2009ApJ...705L.168G}.

The ultraluminous X-ray source  CXO J033831.8-352604 is associated with a globular cluster 
in the Fornax elliptical galaxy NGC 1399, with 
$L_{\rm X}\ga 2 \times 10^{39} {\rm ergs \ s^{-1}}$ \citep{2009arXiv0908.1115I}.
This source shows strong [O III] emission lines (less broad; $\sigma\sim70$ km/s) 
and little or no hydrogen emission.  It may be another black hole-white dwarf X-ray binary accreting near Eddington, 
although \citet{2009arXiv0908.1115I} suggest a tidal disruption of a white dwarf by an intermediate mass black hole (IMBH).

How common are such systems?  
\citet[][K06]{kim06}  found 8 such ULXs,  with $L_{X}\ga 10^{39}\ {\rm ergs\ s^{-1}}$,  in 
6173 globular clusters, where these GCs have an average mass  of $6\times 10^5 M_\odot$.  
\citet[HB08]{2008ApJ...689..983H} found 2 such objects among 3782 globulars 
with total $L_V\sim10^9\ L_\odot$ (total mass $M \sim ~ 3\times10^9\ M_\odot$).  
These surveys suggest that GC ULXs are present at the rate of 
$2.0^{+1.5}_{-1.0}\times 10^{-9}$ and $7^{+15}_{-6}\times 10^{-10}$ per $M_\odot$, 
respectively, using Gehrels' statistics at 90\% confidence \citep{1986ApJ...303..336G}.  
From the overlap in these numbers, we estimate that there is $\sim$1 such object per $1-2\times10^9\ M_\odot$ in GCs.  

These surveys were not completely independent, as several galaxies (including NGC 1399) 
were contained in both surveys.  Some fraction of the observed globular ULXs may not be BH-WD systems, but accreting IMBHs.
The similarities between the two well-observed ULXs, discussed in \cite{2009arXiv0908.1115I}, 
suggest that they are either both BH-WD binaries or both accreting IMBHs.
For this paper, we explore the consequences of the assumption that all globular cluster ULXs 
are BH-WD binaries, in particular whether such numbers can be produced, and by what mechanisms. 

Let us consider now what this frequency implies for the formation rate of such systems per BH.
In globular clusters, roughly each 150-200 $M_\odot$ of currently remaining stellar mass produced a BH in the past,
where half of these BHs have masses above $10 M_\odot$
(here, less cluster mass per BH is required in a metal-poor cluster). 
The retention fraction of BHs immediately after their formation, 
if supernova kicks are taken into account, is about 30-40\%, for the escape velocity of 50 km/s \citep{2006ApJ...650..303B}.
We estimate therefore that, if no dynamical interactions between stars occur, a present day average cluster in the 
\cite{kim06} sample ($6\times 10^5 M_\odot$) would retain about 500-800 BHs with masses above $10 M_\odot$.

Standard consideration of the consequences of dynamical interactions between BHs and other stars 
in clusters assumes that there is a tendency towards equipartition, where BHs quickly detach from other stars 
and form a BH subcluster  \citep[this is also known as the Spitzer instability, ][]{1969ApJ...158L.139S}.
Within this subcluster they interact only with each other and, as a result, fairly quickly evaporate from the cluster.
Assuming the evaporation times of \citet{2000ApJ...539..331W} and \citet{kalogera_04}, only one BH could remain per cluster at the current epoch.
This evaporation timescale assumed simple evaporation from an isolated cluster  and is probably on the high side.
Detailed numerical calculations of BH subcluster dynamical evolution, however, have shown that in some circumstance a significant fraction
of BHs will remain at 10 Gyr time -- in massive clusters, the fraction of BHs remaining in the cluster at 10 Gyr is $\sim 20\%$ of the BHs
retained after their formation \citep{oleary_06}. 
It has been also noticed that in massive clusters the BH subcluster does not reach equipartition at Hubble time. 
Similar results were found in other studies \citep{2009arXiv0910.0546D}, where a whole globular cluster, including
normal stars, was modeled using a Monte Carlo treatment. 
Even though black holes were found to be strongly centrally concentrated and interacting mainly with each other, a significant fraction of 
initial BHs (up to 25\%) not only remained in the cluster but also participated in interaction with other stars.
(It is not clear which fraction is purely in the BH subcluster and which fraction did not detach from the rest of
stellar population.) 
We assume therefore that the number of massive BHs ($\ga 10 M_\odot$) retained, and available for interactions with core stars, could be from one per average cluster in \cite{kim06} sample (i.e. roughly 4000-6000 BHs in each whole observed sample, K06 and HB08)  
up to 125-200 per average K06 cluster or 160-260 per average HB08 cluster 
(i.e. about $5\times 10^5 - 1.2\times 10^6$ BHs in the observed samples).
The upper limit corresponds to the case when the number of massive BHs is 10\% of all initally formed massive BHs.

The lifetime of the persistent mass transfer (MT) at the Eddington level ($\sim 4\times 10^{39}\ {\rm ergs\ s^{-1}}$) from a WD to such a BH, 
once started, is $\tau_{\rm X}\approx 2\times 10^5$ yr and is about 3 times longer for
ULX luminosities  $L_{X}\ga 10^{39}\ {\rm ergs\ s^{-1}}$.
If all but one BH is evaporated, the formation rate of BH-WD binaries must be
extremely high -- up to one X-ray binary formation per BH per Gyr.
The minimum required formation rate, when a good fraction of BHs is retained which also do not 
form a BH subcluster, is about $4\times 10^{-3}$ per BH per Gyr (and more likely $10^{-2}$ per BH per Gyr, if many 
clusters are metal-rich and have lower escape velocity than 50 km/s). 
We therefore explore which dynamical formation channel would be able
to provide the formation of BH-WD X-ray binaries at rates of $4\times 10^{-3}$ to 1 per BH per Gyr in an average dense massive cluster.
We can also define this as  $4\times 10^{-3}/f_{\rm BH,0.1}$   per BH per Gyr, where $f_{\rm BH,0.1}=0.1 f_{\rm BH,tot}$ is the fraction of BHs
that is retained in the cluster and not detached in the BH subcluster, and is normalized to 10\%
of all initially formed massive BHs ($\ga 10 M_\odot$). 
A smaller rate is possible only if essentially all globular clusters have experienced significant tidal stripping and
the number of remaining BHs per present stellar mass unit is significantly higher than 1 BH per 150 $M_\odot$.
To analyze the  dynamical formation, we will proceed in reverse order: first, we will consider which BH-WD binaries, 
once formed, can become X-ray binaries, and
then we will consider at what rates these BH-WD binaries can be formed via different dynamical channels.

\section{Fate of a BH-WD binary}

\subsection{Characteristic times}

\label{bsfate}

In a dense stellar system, once a BH-WD binary is formed through a dynamical encounter, 
its further binary evolution could be affected by subsequent encounters with other stars. 
The cross section for an encounter between two objects of total mass
$m_{\rm tot}$
with a distance of closest approach less than $r_{\rm max}$
is computed as
\begin{equation}
\sigma = \pi r_{\rm max}^2 \left ( 1 + \frac{v_{\rm p}^2}{v_{\infty}^2}\right),
\end{equation}
where the second term accounts for gravitational focusing, with
$v_{\rm p}^2=2Gm_{\rm tot}/r_{\rm  max}$ 
and $v_\infty$ the relative velocity at infinity.
For strong interactions, $r_{\rm max}$ is usually on the order of the
binary semimajor axis: $r_{\rm max}=k a$, where $k$ is of order unity \citep{hutbah83}.
In the limit of strong gravitational focusing, $v_{\rm p}^2 \gg
v_{\infty}^2$ and
\begin{equation}
\sigma = 2 \pi G k a m_{\rm tot} v_\infty^{-2}. \label{sigma_equation}
\end{equation}

\begin{figure}
 \includegraphics[scale=.43]{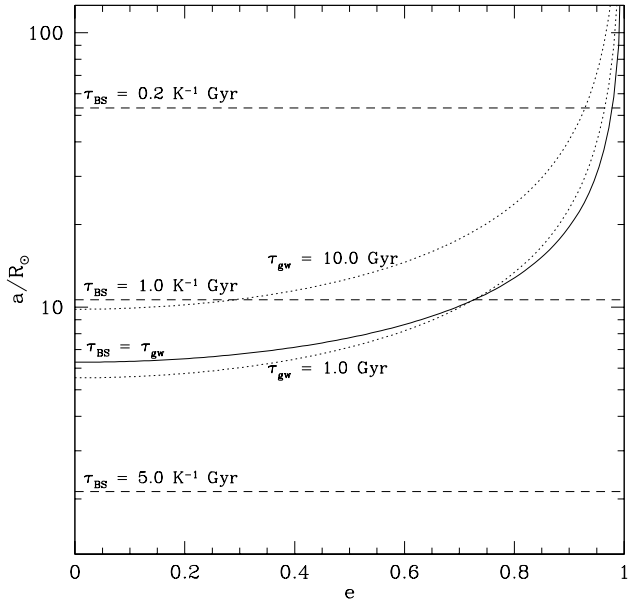}
 \caption{Orbital parameters (initial separation $a/R_{\odot}$ and eccentricity $e$) 
of a 15 $M_{\odot}$ BH and 0.6 $M_{\odot}$ WD binary system. 
Lines of constant merger time $\tau_{\rm gw}$ (dotted curves) and
encounter time $\tau_{\rm BS}$ with a single star (dashed lines) are shown at specified values. 
Also shown is the separation curve $a_{\rm sep}$ for which
$\tau_{\rm gw}$ = $\tau_{\rm BS}$ (solid curve), assuming the
dimensionless parameter $K$ defined by equation (\ref{binsin}) equals 1.
\label{tau}}
\end{figure}

In our case, the first object is the BH-WD binary of mass
$m_{\rm BHWD}=m_{\rm BH}+m_{\rm WD}$ while the second object is a core
star of mass $m_\star$.
In globular clusters, the close approaches that we are interested in have
$v_{\rm p}^2 \gg v_{\infty}^2$, and the BH mass is significantly more massive
than both its WD companion and a typical core star.  Thus, the rate at which a
BH-WD binary undergoes a strong (binary-single) encounter is
\begin{eqnarray}
\label{binsin}
\Gamma_{\rm BS} & = & \sigma n_{\rm c} v_\infty \\
 &\simeq & 2 \pi G  k m_{\rm BH} n_{\rm c} a v_{\infty}^{-1} \nonumber
 \\
&\simeq & 0.1k 
\frac{m_{\rm BH}}{15 M_\odot} \frac{n_{\rm c}}{10^5 {\rm pc}^{-3}}
\frac{10{\rm km\, s}^{-1}}{v_{\infty}} \frac{a}{R_\odot}\, {\rm per \
  Gyr} \nonumber \\
&= & 0.1 K \frac{a}{R_\odot}\, {\rm per\ Gyr}. \nonumber
\end{eqnarray}
The final equal sign in equation (\ref{binsin}) defines the
dimensionless parameter $K$.
The time-scale for a BH-WD binary to experience a strong encounter can
be calculated as $\tau_{\rm BS} = 1 / \Gamma_{\rm BS} =
10.6\,K^{-1} R_\odot/a$ Gyr: see Figure \ref{tau}.
Throughout this paper, we consider
clusters with
core number densities $n_c$ near
$10^5$ pc$^{-3}$, velocity dispersions of $\sim 10$ km s$^{-1}$, and black hole
masses of $\sim15M_\odot$: consequently, $K$ is of order unity.

Once formed, the orbit of a BH-WD binary starts to shrink due to gravitational radiation.
The time $\tau_{\rm gw}$ for the orbit to decay enough for mass
transfer to commence depends on the initial post-encounter binary separation $a$ and eccentricity $e$. 
Applying \cite{peters64} equations, we can find the maximum semimajor
axis $a$, as a function of post-encounter eccentricity $e$, 
for which a system will start MT within any specified time, e.g., 1
Gyr and 10 Gyr (see the dotted curves in Fig.\ref{tau}).

In Figure \ref{tau}, we also show the semimajor axis $a_{\rm sep}(e)$
such that $\tau_{\rm BS} = \tau_{\rm gw}(e)$, assuming $K=1$.
BH-WD binaries with post-encounter semimajor axes $a(e)<a_{\rm sep}(e)$  
will become BH-WD X-ray binaries.
BH-WD binaries with post-encounter semimajor axes $a(e)>a_{\rm sep}(e)$  
will have one or more strong encounters before they shrink to the
point of merger due to gravitational waves radiation.  In the next
subsection, we consider the possible outcomes of such encounters.

\subsection{Encounters with single stars}

Here we consider the outcomes and their frequencies for encounters between a BH-WD and a single star.
An encounter between a soft binary and a third star can lead
to ionization if the third star approaches with sufficient speed;
specifically, for ionization to be energetically possible, the
relative velocity at infinity
$v_\infty$ between the binary and the third star must exceed the
binary's critical velocity $v_{\rm c}$ defined such that  the total energy of the binary-single system  is zero:
\begin{equation}
v_{\rm c}^2 = \frac{m_{\rm BHWD}+m_\star}{m_{\rm BHWD}} \frac{m_{\rm WD}}{ m_{\star}}\frac{G m_{\rm BH}}{ a}.
\end{equation}
For $m_{\rm BH}=15 M_\odot$, $m_{\rm WD}=0.6M_\odot$, and $m_\star=0.6 M_\odot$, the critical
velocity $v_c$ is less than 10 km s$^{-1}$
only if the semimajor axis $a$ of the binary exceeds $\sim3\times10^4 R_\odot$.
Even for encounters between a BH-WD binary and a 15$M_\odot$ black hole, the 
critical velocity $v_c$ is less than 10 km s$^{-1}$ only for $a\ga 2000R_\odot$.
Consequently, in a typical cluster with velocity dispersion $\sim 10$km
s$^{-1}$, ionization by itself can {\it  not} destroy BH-WD binaries that are able 
to start MT within 10 Gyr, as all such binaries are easily of sufficiently
small semimajor axis (see Fig.~\ref{tau}). Essentially, a BH-WD binary with 
$a\la 2000 R_\odot$ can be conservatively classified as a hard binary.

A strong encounter between a hard binary and a third star would
ultimately lead to preservation of the BH-WD binary, a companion exchange, or a physical collision.
Such outcomes, which we now consider, could potentially prevent a
BH-WD binary from ever reaching the MT stage.
We note that if an encounter occurs with another binary, a possible outcome is also a hierarchically stable triple system;
we consider this option in the next subsection.

The collision cross section for binary-single interactions 
is approximately (for a non-eccentric binary)\footnote{This
  approximation was derived using results from \cite{freg04}.}:

\begin{equation}
\sigma_{\rm coll} \sim \pi a^2 \left(\frac{V_{\rm g}}{v_\infty}\right)^{2} \left(\frac{215 R}{a}\right)^{0.65},
\label{tcoll}
\end{equation}
where $V_{\rm g}^2=Gm_{\rm tot}/(2a)$ and the maximum radius $R$ among the
encounter participants is presumed to be less than $a$.  Our definition of $V_g$ follows that of
\cite{1996ApJ...467..359H}.  We note that $V_{\rm g}=v_{\rm c}$ in the case of three
equal mass stars.   
We have verified with numerical experiments for
  many mass combinations that the use of $V_{\rm g}^2$ yields the
  appropriate scaling of the cross section with mass (provided
$v_\infty \ll V_{\rm g}$).

Using equations (\ref{sigma_equation}) and (\ref{tcoll}), we can now estimate the fraction of binary-single encounters that result in a collision:

\begin{equation}
\frac{\sigma_{\rm coll}}{\sigma} = \frac{1}{4k} \left(\frac{215 R}{a}\right)^{0.65},
\label{collfrac}
\end{equation}
where $k$ is the ratio of the maximum pericenter considered as an
encounter to the semimajor axis $a$ of the binary.
Equation (\ref{collfrac}) implies that if a non-eccentric BH-WD has $a\la 5 R_\odot$,
then all encounters with $R=0.6R_\odot$ MS star at $k<2$ will result in a merger, while
for encounters with $k\le1$,  $a\la 15 R_\odot$ will result in a merger.
Consequently, we expect that most BH-WD binaries that are able 
to start MT within 10 Gyr (see Fig.~\ref{tau})
will experience a merger if a binary-single encounter with a MS star occurs.

To examine the collision cross section more carefully, we
perform scattering experiments as in \cite{freg04}, but for our BH-WD binary with the separations from 1 to 100 $R_\odot$
and fixed $v_\infty=10$ km/s.  We find the $k_{\rm coll}$, defined by equation
(\ref{sigma_equation}) with $\sigma=\sigma_{\rm coll}$ and $k=k_{\rm coll}$, varies from $\sim1$ for $a=1R_\odot $ to $\sim0.1$ for $a=100 R_\odot$.  For
example, $k_{\rm coll}\approx 0.24$ for $a=10 R_\odot$, and, accordingly from
equation (\ref{binsin}), the merger rate for such BH-WD binaries that are able 
to start MT within 10 Gyr exceeds 0.2 per Gyr.
We note that effectively this rate could be a factor of few smaller as   a fraction
of collisions will result in a merger between a BH and a MS star.
We have verifed with hydrodynamicsl simulations that at least some of these collisions
will not affect strongly the initial BH-WD binary.

The relative fraction of WDs in the core, at the age of several to 14 Gyr, is  $f_{\rm WD}\sim 0.2$ of the total core population, 
and $\Gamma_{\rm BS,WD}=f_{\rm WD}\Gamma_{\rm BS}$. We also note that some simulations have shown that WDs can contribute up to 70\%
of the total core population, as was found in simulations performed for \cite{2009ApJ...695L..20F} without WD birth kicks.
From eq.\ (\ref{collfrac}), we roughly estimate that a physical collision with a
$R=0.01R_\odot$ WD would occur only in binaries with  $a\la0.1 R_\odot$,
which is well below $a_{\rm sep}(e)$.
A more likely outcome in such cases would be preservation or a companion exchange.
Exchanges preferentially occur if the incoming star is more massive than pre-encounter companion and accordingly could happen
if BH-WD binary had a light WD companion. As a result of exchange, the post-encounter binary separation will
be {\it increased} by the ratio of new companion mass to the old companion mass. 
To reiterate, we consider here only the cases when a BH-WD binary has $a(e)>a_{\rm sep}(e)$.
Therefore, the resulting binary will most likely never be able to reach MT, 
as its binary separation will exceed the value necessary for MT to start within 10 Gyr. 
If a binary is preserved, though, it will most likely then experience a consequent 
encounter with a single star. Even if WDs make up as much as 70\% of
the core population,
this consequent encounter has high probability to have a MS star as a participant and
therefore to result in a merger.
We conclude that strong encounters with single stars will not create BH-WD binaries
with $a(e)<a_{\rm sep}(e)$ and therefore will not lead to a direct formation of a BH-WD X-ray binary.

\subsection{Encounter with binary stars: role of triples}

\begin{figure}
 \includegraphics[scale=.4]{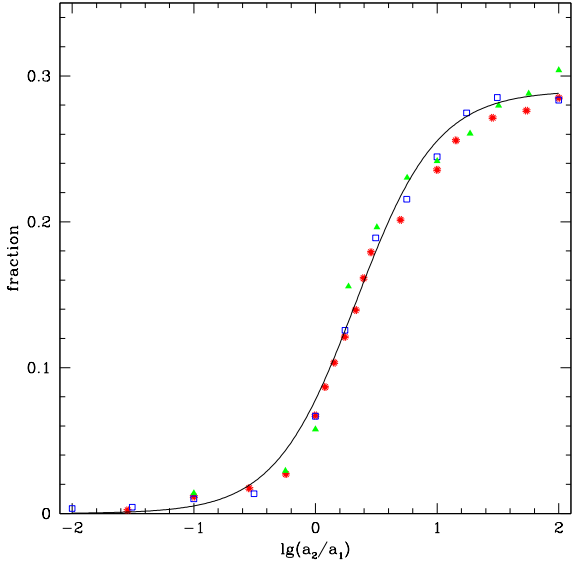}
 \caption{ Fraction of encounters, for $k=20$, that result in a triple formation. 
Encounters are between a BH-WD $15 M_\odot + 0.6 M_\odot$ binary with separations 
$a_1=15, 35,$ and $80 R_\odot$ (green triangles, red stars and blue squares accordingly)  
and $0.6M_\odot+0.6M_\odot$ binary with separations $a_2$.
Each point shows the result for 10 000 encounters.
The solid line is the fitting  with $0.29 \left (1+e^{(-3\lg(a_2/a_1)+1)} \right )^{-1}$
\label{trip_form}}
\end{figure}

If a BH-WD has an encounter with another binary, a possible outcome is a hierarchically stable triple.
For two identical binaries, the maximum impact
parameter for which strong encounters still could occur is a bit larger
than for a binary-single encounter \citep{mikkola83}:
\begin{equation}
\sigma_{\rm BB} \approx 4.6 \pi (a_1+a_2)^2 \frac{v_{\rm c}^2} {v_\infty^2}.
\label{ktrip}
\end{equation}
Here $v_{\rm c}$ is the critical velocity of the two binaries
with the masses $m_1$ and $m_2$:
\begin{equation}
v_{\rm c}^2 = \frac{G}{\mu} \left ( \frac{m_{11}m_{12}}{a_1} + \frac{m_{21}m_{22}}{a_2} \right ),
\end{equation}
where the reduced mass $\mu = m_1 m_2 / (m_1+m_2)$, where
$m_{11}$, $m_{12}$, $m_{21}$ and $m_{22}$ are the masses of the
binary components, and where $a_1$ and $a_2$ are the semimajor axes.

For hard binaries with equal masses and binary separations, $\sim 25\%$ of 
all binary-binary encounters within maximum approach described by eq.\
(\ref{ktrip})
and ${v_{\rm c}^2}/ {v_\infty^2}=10$ form a triple system \citep{mikkola83, freg04},
with triple formation increasing with  ${v_{\rm c}^2}/ {v_\infty^2}$.
For a larger mass ratio between binaries, or for a softer incoming binary of less mass, this fraction can be even higher.

We performed numerical experiments in which BH-WD binaries
consisting of a $15 M_\odot$ BH and a $0.6 M_\odot$ WD 
encountered at $v_\infty=10$km/s binaries consisting of 0.6 $M_\odot$ companions.
We record the fraction of encounters  that result in a stable triple
with an inner binary
consisting of the initial BH and a companion $0.6 M_\odot$ star.
We run the simulations for 3 values of BH-WD binary separation and
11-17 values of the incoming binary separations, performing 10000 encounters for each.
For the numerical integration, we use {\tt Fewbody} \citep{freg04}, a small $N$-body integrator.
The results demonstrate that periastrons somewhat larger than those suggested by eq.(\ref{ktrip})
can result in a triple formation, and we choose the maximum possible periastron for the encounters to be fairly 
large, $r_{\rm max, p}=20(a_1+a_2)$ (that is, $k=20$). The impact parameters for every encounter are distributed uniformly
in area. The eccentricities are distributed thermally.
The results are shown at the Fig.\ref{trip_form}.

We find, that for almost all the encounters in which the incoming
binary is wider than the BH-WD binary, the efficiency
of triple formation is  $\sim 20-30\%$ with $k=20$. This gives 

\begin{equation}
\sigma_{\rm triples} \approx 120 \pi (a_1 + a_2)^2 \left ( 1 + \frac{G m_{\rm BH}}{10 (a_1+a_2) v_\infty^2} \right ).
\end{equation}
The  formation rate of triple systems, for hard binaries,  
compared to to binary-single encounters rate (\ref{binsin}) with $k=2$, is then

\begin{eqnarray}
 \frac{\Gamma_{\rm triples}(a_1, a_2)}{\Gamma_{\rm BS}(a_1)} &\sim&  3\sqrt{2} f_{\rm wb}  \left ( 1+\frac{a_2}{a_1} \right ).
\end{eqnarray}
Here  $f_{\rm wb}$ is the fraction of stars that are hard binaries,
but still wider than the BH-WD binary.  In deriving the above equation, we have applied equipartition of energy, so that the $v_\infty$  of a typical binary in the core is $\sqrt{2}$ times less than that of a typical single star.

The overall binary fraction in observed dense, but not core-collapsed, clusters at the current epoch, 
is about a few per cent \citep{Albrow01,bf_nata,Milone08}.  
With $f_{\rm wb}=5\%$ and a flat distribution of binary separations between, e.g., 20 and 
$2000 R_\odot$,  a BH-WD binary that has $a=20 R_\odot$ can form a triple about 30 times per Gyr.
This exceeds the interaction rate with single stars.

Although these triples are stable in isolation, they can be destroyed during their next dynamical encounter.
Before this next encounter occurs,  the presence of the third component can significantly affect the eccentricity of the inner BH-WD binary via the Kozai mechanism \citep{kozai_62}, 
provided the inclination of the outer orbit is large enough
($\sin i_0> (2/5)^{1/2}$, or equivalently $i_0 \ga 39{\degr}$, the Kozai angle). 
The maximum eccentricity that can be achieved via the Kozai mechanism is \citep[e.g.,][]{Innanen97_kozai, Eggleton01_triples}

\begin{equation}
e_{\rm max} \simeq \sqrt{5/3 \sin^2(i_0)-2/3} \ .
\end{equation}

In dynamically formed triples, the inclinations are distributed almost uniformly,
and so the number of Kozai triples is expected to be proportional 
to the solid angle \citep[e.g.,][ where, depending on the sample of runs, 
the triples affected by the Kozai mechanism constituted from 30\% to 40\%
of all formed triples, while a uniform distribution predicts 37\%]{natatrip08}.
Thus, the fraction of the dynamically formed triples with a
post-encounter inclination greater than $i$ is $f_i=1-\sin i$.

The BH-WD binaries that could become MT systems are only those that have  
$a$ less than the maximum $a_{\rm sep}$ ($\la 80 R_\odot$ for $e < 0.99$).
To find the formation rate of such triples, we first determine the minimum value of $e_{\rm max}$ such that a triple 
with the specific binary separation $a$ must achieve: this is found 
by inverting the numerically obtained $a_{\rm sep}(e)$ at $k=2$.
The obtained required $e_{\rm max}$ then provides the fraction of
formed triples that can be bring the inner binary into MT via the Kozai mechanism:
\begin{equation}
f_{i} = 1-\sin i = 1-\sqrt{3/5\ e_{\rm max}^2+2/5}.
\label{fi}
\end{equation}

We define triple induced mass transfer (TIMT) systems to be those
binaries that are brought into MT via the the Kozai meachanism.
We find that the fraction of TIMT systems ranges between 0.07 (for $a=15 R_\odot$) and 0.01 (for most separations, $a\ga 40 R_\odot$).
Combining it with the triple formation rate, we find that all BH-WD binaries with 
$15 R_\odot \la a_1 \la 80 R_\odot$ could make it to the MT through the triple formation mechanism.
At a conservative level, assuming even that all the binaries that have a binding energy 10 times more
than the kinetic energy of an average object in the core  are destroyed,
we find that a BH-WD binary with $a_1\ga 50 R_\odot$  has a 10-40\% chance to become become a TIMT system 
within 1 Gyr (10\% for $a_1\sim 80 R_\odot$ and 40\% for $a_1\sim 50 R_\odot$); 
this chance is 100\% for  $a_1\la 35 R_\odot$. During several Gyr, all BH-WD binaries with 
$a_1 \la 80 R_\odot$ can become a TIMT system  at least once.

This TIMT system formation will be successful only 
if the time between subsequent encounters is longer than the time necessary to achieve $e_{\rm max}$.
The period of the cycle to achieve  $e_{\rm max}$ is  \citep{Innanen97_kozai,Miller02_kozai}

\begin{equation}
\tau_{\rm Koz}  \simeq  \frac{0.42 \ln (1/e_{\rm i}) }{\sqrt{\sin^2(i_0) -0.4}}  
\left ( \frac{m_{i1}+m_{i2}}{m_{\rm o}} \frac{b^3_{\rm o}}{a_1^3} \right ) ^{1/2} 
\left ( \frac{b^3_{\rm o}}{ G m_{\rm o}} \right )^{1/2} ,
\label{tkoz}
\end{equation}

\noindent where $m_{i1}$ and $m_{i2}$ are the companion mases of the inner binary,
$e_{\rm i}$ is the initial eccentricity of the inner binary,
$m_o$ is the mass of the outer star,
$a_{\rm o}$ is the initial semimajor axis for the outer orbit,
and $b_{\rm o}= a_{\rm o}(1 - e_{\rm o}^2)^{1/2}$ is the semiminor axis of the outer orbit.
Roughly, $a_{\rm o}$ will be of the order of magnitude of $a_2$.  Considering the
extreme case $a_2\gg a_1$, for which $\tau_{\rm Koz}$ would be maximum, 
we find:

\begin{eqnarray}
\frac{\tau_{\rm Koz}}{\tau_{\rm BB}} &\approx&  4\times 10^{-14} f_{\rm b} \frac{0.42 \ln (1/e_{\rm i}) }{\sqrt{\sin^2(i_0) -0.4}}  
\left ( \frac{a_{\rm 2}^2}{a_1 R_\odot} \right ) ^{5/2}   \\
& \ & \left ( \frac{m_{\rm BH}}{15 M_{\odot}}\right)^{3/2} \frac{M_{\odot}}{m_{\rm o}}  
\left( 1 - e_{\rm o}^2 \right )^{3/2} 
 \frac{n_{\rm c}}{10^5 {\rm pc}^{-3}} \frac{10{\rm km/s}}{v_{\infty}}  \nonumber \\
\frac{\tau_{\rm Koz}}{\tau_{\rm BS}} &\approx&  4\times 10^{-14} \frac{0.42 \ln (1/e_{\rm i}) }{\sqrt{\sin^2(i_0) -0.4}}  
\left ( \frac{a_{\rm 2}}{R_\odot} \right ) ^{3} \left ( \frac{R_\odot}{a_1} \right ) ^{1/2}   \\
& \ & \left ( \frac{m_{\rm BH}}{15 M_{\odot}}\right)^{3/2} \frac{M_{\odot}}{m_{\rm o}}  
\left( 1 - e_{\rm o}^2 \right )^{3/2} 
 \frac{n_{\rm c}}{10^5 {\rm pc}^{-3}} \frac{10{\rm km/s}}{v_{\infty}}  \nonumber
\label{taukozcol}
\end{eqnarray}
Here $f_b$ is the binary fraction,
$\tau_{\rm BB}=\Gamma_{\rm BB}^{-1}=(\sigma_{\rm BB}f_{\rm b}n_{\rm c}v_\infty)^{-1}$, and $\tau_{\rm BS}=1/\Gamma_{\rm BS}$ has been evaluated at $k=2$.

An average dynamically formed triple would have $e_{\rm o}\approx 0.9 $ \cite{natatrip}. For the purpose of the estimate, 
we adopt that a BH-WD binary would have its initial eccentricity distributed thermally, with an average $e_{\rm i} \sim 2/3$ 
and adopt $i_0=54^o$ (this is the mean inclination of dynamically formed Kozai affected triples).
From eq.(\ref{taukozcol}), we have then that triples with $15 R_\odot \la a_1 \la 80 R_\odot$ 
and $a_{\rm  2} \la 10^5 R_\odot$ (where $4500 R_\odot$ is the boundary between hard and soft binaries) 
would have their Kozai time significantly shorter than their collision time with either binary or single stars.
We conclude that once a potential TIMT system is formed, it will succeed in bringing its inner BH-WD to the mass transfer before
its next encounter. Accordingly, all BH-WD binaries with $a\la 80 R_\odot$ will become X-ray binaries via TIMT mechanism.

\label{kozaitrip}

\subsection{Multiple encounters: hardening}

It is well known that through multiple fly-by encounters hard binaries get harder \citep[e.g.,][]{hut83}.
The number of encounters necessary to change the binary energy, e.g., by a factor of four, can be estimated as
\citep{heggiehut_03}

\begin{equation}
N_{\rm hard} \approx \log(4)/\log(1+\delta),
\end{equation}
where $\delta$ is the relative change in the binary's binding energy caused by  an encounter.
This change, for an eccentric binary with a large mass ratio of one star to a companion and an encountering  
$m_3$ star, if occurred at the distance $q$ ($q\gg a$),
can be roughly estimated as \citep[e.g., ][]{heggie75,2003CeMDA..87..411R}

\begin{equation}
\delta \approx \frac{m_3}{m_{\rm BH}} \left( {\frac{a}{q}} \right)^{3/4} e^{-(q/a)^{3/4}}.
\end{equation}
Assuming that all hardening happened with the maximum $\delta\sim {m_3}/{m_{\rm bh}}\sim0.04$, it can therefore be estimated that,  the hardening of a 1000 $R_\odot$ BH-WD binary to, e.g., $35 R_\odot$ should take about 100 encounters. As the collision time for a 1000 $R_\odot$ BH-WD binary is only about $10^7$ yr,
and is still only about $10^8$ yr for a 35 $R_\odot$ BH-WD binary, it is plausible therefore  that 
all 1000 $R_\odot$ BH-WD binary can be hardened to $35 R_\odot$ within
a few Gyr. 
We choose here $35 R_\odot$ as the important separation
at which triple formation starts to have 100\% efficiency in making this binary a MT BH-WD binary (see \S\ref{kozaitrip}).

However, this is an idealistic picture. 
There are plenty of encounters that this binary would have undergone in order to reach 35 $R_\odot$, 
and certainly not all them will be simple fly-by hardening encounters.
Some of the encounters would result in exchanges (where the lighter companion is often exchanged with a more massive incoming star,
and the binary gets wider), mergers, or even binary ionizations, reducing therefore the chance for a binary to harden
to the separation we are interested in.

We set up a numerical experiment, by taking a BH-WD binary with the specific initial binary separation $a_{\rm i}$. 
At the collision rate predicted by the current binary separation and the adopted $n_{\rm c}=10^5 \ {\rm pc^{-3}}$,
we bombard the binary with single stars drawn from the simplified core's population, randomizing impact parameters
and using {\tt Fewbody}.
Initial eccentricities of BH-WD binaries for this experiment is
adopted to have a thermal distribution.
A simplified core population is taken from Monte Carlo runs \citep{ivanova08} and has only 4 different groups for the non-BH population: 
stars with masses $0.22, 0.506, 0.75$, and $1.13 M_\odot$, with according $33\%, 38\%, 25\%$, and 
$4\%$ contributions to the non-BH population. 
Then we vary the BH component, starting from the highest possible, 
where the number of BHs is 0.4\% of all other stars in the core
(this is about 100\% of the initially formed BH population, see the discussion in \S1, or  $f_{\rm BH,0.1}=10$), 
0.04\% (this corresponds to the case of $f_{\rm BH,0.1}=1$), 
0.004\%, and no contribution at all (this correspond to the case when only one BH, which is in the considered BH-WD binary, is left).
Here, one BH corresponds to $0.001\%$ of the core population.

We consider the fraction of BH-WD binaries that would harden to $35
R_\odot$ within 10 Gyr as well as the
average time it will take them. 
We also separately consider cases where we insisted that the {\it original} BH-WD binary
would made it to $35 R_\odot$, or simply {\it any} binary containing initial BH would made it (allowing multiple exchanges). 
The latter case could correspond to the case when almost all the stellar non-BH population in the core are WDs \citep[e.g.,
in some runs performed in][WDs fraction in the core reached 70\%]{2009ApJ...695L..20F}.
The results are shown in the Table \ref{table_hard}.

\begin{table}
\caption{Hardening}
\begin{tabular}{| l  r  r  r   r   r   r   r   r |}
\hline 
 & \multicolumn{4}{c}{Hardened fraction [per cent]} & \multicolumn{4}{c|}{Average time [Gyr]} \\
 & \multicolumn{8}{c|}{Black hole fraction in the total core population [per cent]} \\
 & 0.4 & 0.04 & 0.004 & 0 & 0.4 & 0.04 & 0.004 & 0 \\
 & \multicolumn{8}{c|}{Corresponding $f_{\rm BH,0.1}$} \\
 & 10 & 1 & 0.1 & 0 & 10 & 1 & 0.1 & 0 \\
$a_{\rm i}$ &  \multicolumn{8}{c|}{ }\\
\hline
\multicolumn{9}{|c|}{Hardening to 35 $R_\odot$} \\
\multicolumn{9}{|c|}{Original binary survival only} \\
$100 R_\odot$ & 4.6 & 12.9 & 14.8 & 14.9  & 0.86  & 1.38 & 1.48 & 1.52 \\
$250 R_\odot$ & 0.53 & 2.5 & 3.4 & 3.4 &1.07 & 1.9 & 1.9 & 1.9 \\
$500 R_\odot$  & 0. & 0.8 & 1.0 & 1.3 & - & 1.9 & 2.1 & 2.3\\
\multicolumn{9}{|c|}{Exchanges are allowed} \\
$100 R_\odot$  & 9.4 & 30.4 & 38.1 & 38.9 & 1.06 & 1.50 & 1.58 & 1.60 \\
$250 R_\odot$  & 1.3 & 9.8 & 13.9 & 13.3 & 1.17 & 1.85 & 2.00 & 2.07 \\
$500 R_\odot$ & 0. & 3.2 & 4.8 & 5.9  & -  & 1.78 & 2.15 & 2.09\\
\multicolumn{9}{|c|}{Hardening to 80 $R_\odot$} \\
\multicolumn{9}{|c|}{Original binary survival only} \\
$100 R_\odot$ & 37.9 & 56.0 & 58.2 & 60.6  &  0.22  & 0.29 & 0.30  &  0.31 \\
$150 R_\odot$ &  12.7&  27.5 & 30.0  &  30.9  & 0.35  & 0.52  & 0.57 & 0.57 \\
$200 R_\odot$ &  6.5 & 16.9 & 19.1 &  20.7 &   0.41 & 0.64 & 0.68 &  0.70 \\
$250 R_\odot$ &  3.5 & 11.3 & 14.4 & 14.4  &  0.46 &  0.73 & 0.77 &  0. 77 \\
$500 R_\odot$ & 1.1 & 3.1 & 3.8 & 3.8  & 0.37  &  0.90 & 0.98 & 0.98 \\
\hline 
\end{tabular}
\label{table_hard}
\footnotesize{The table shows what fraction of BH-WD binaries with $15 M_\odot$ BH  and $0.6 M_\odot WD$ and initial binary separations $a_{\rm i}$
can be hardened to 35 $R_\odot$ and 80 $R_\odot$, as a function of the relative BH content in the core: 0.4\%, 0.04\%, 0.004\% of the population,
or no any other BHs present ($f_{\rm BH,0.1}$ are the corresponding values of the fraction of BHs that 
are available for interactions with normal stellar population, see \S1).  The average time corresponds to the time
that it takes to harden those binaries that successfully got harder.
The number of encounters used for the statistics: 10000 for $100 R_\odot$, 5000 for $100 R_\odot$ and 1000 for $500 R_\odot$.}
\end{table}

There are several trends in the results, and all of them well expected theoretically.
Indeed, as expected, the fraction of survived and hardened BH-WD binaries  $f_{\rm hard}$  is well below 1.
The fraction $f_{\rm hard}$ decreases with increasing $a_{\rm i}$ and increases
when the BH population in the core drops, as other BHs contribute significantly into exchange reactions or ionisation. 
The average time that it takes to harden a binary agrees with the analytic estimate above and is about 1-2 Gyr.
Even though $f_{\rm hard}$ increases when the BH population drops,
it does not increase by a factor comparable to the relative drop in the BH population. Accordingly, since
the resulting formation rate of hardened BH-WD binaries (their production from initially wider binaries) 
$\Gamma_{\rm hard}\propto f_{\rm BH,0.1} f_{\rm hard}$, the highest $\Gamma_{\rm hard}$ occurs 
in an unlikely case when most of initially formed BHs are retained ($f_{\rm BH,0.1}=10$).
The fraction $f_{\rm hard}$ for $f_{\rm BH,0.1}\la 1$ does not vary much with $f_{\rm BH,0.1}$.

We also study what fractions can be hardened 
to $a=80 R_\odot$, for our optimistic scenario when all BH-WD binaries  with $a=80 R_\odot$ make it to the MT via TIMT.
In this case, siginificant fractions of wide binaries can be sucesfully hardened.

\section{Formation of a BH-WD binary}

In the previous Section we considered what can happen to a BH-WD binary in a dense stellar system. Here we analyze
which kind of a BH-WD binary could be formed and, accordingly, which formation mechanism dominates in the formation
of potential BH-WD X-ray binaries. We neglect the possibility  that an unperturbed primordial binary with a BH could be formed
and survive in a dense stellar system (e.g., \cite{2009arXiv0910.0546D} has shown that the 
fraction of primordial BH binaries is well below 1\% compared to all BH binaries). 
It has been discussed that there are two processes through which 
a BH could acquire a non-degenerate stellar companion: exchange interactions with (primordial) binaries and tidal captures  
\citep[for a thorough discussion of the latter mechanism with a BH and references therein, see][]{kalogera_04}.
Here, we do not consider tidal captures as they are inefficient for interactions between a BH and a WD;
it was suggested that these encounters would result rather in a tidal disruption of a WD or its nuclear ignition  \citep{2009ApJ...695..404R}.
We consider exchange interactions, when an acquired companion is a
WD, as well as two other possible mechanisms:
physical collision with red giants (RGs) and three-body binary formation. The latter mechanism is usually neglected from consideration,
as for all non-degenerate stars it is assumed that such encounters would rather lead to a merger than a formation of a hard binary.

\subsection{Exchange encounters}

In a binary that has been newly formed via exchange, the post-encounter binary separation will be $a_{\rm post}\sim a_{\rm pre} m_{\rm BH}/m_{\rm c}$ where $m_{\rm c}$
is the lower mass companion replaced by a BH. The mass $m_{\rm c}$ would normally not exceed that of an average star in 
the core, and accordingly the post-encounter binary separation will be at least 25 times larger than pre-encounter 
binary separation. In exchange encounters, post-encounter eccentricities are distributed thermally with the mean $e\sim 2/3$. 

From Fig.~\ref{tau}, we see that only those binaries with pre-encounter binary separation $a_{\rm pre}\la 0.3 R_\odot$
could create a binary that will evolve to MT directly (though up to $a_{\rm pre} \approx 3.1 R_\odot$ if the final 
eccentricity is as high as 0.99),  and with $a_{\rm pre}\la 1.4R_\odot$ -- when a formed binary will evolve
to MT through a consequent triple formation. 
In both cases, $a_{\rm pre}$ is small enough to expect a merger if at least one companion
in the pre-encounter binary was a MS star. 

Therefore, to create a BH-WD binary with  $a_{\rm post} \la 35 R_\odot$,
only encounters with WD-WD  binaries are relevant. The fraction of WD-WD binaries is just a few percent 
of the total binary population.
From the simulations set in \cite{ivanova08, nata_wd_06} we find the fraction of the short period WD-WD binaries at 10 Gyr in 
their ``standard" globular cluster (corresponds to our typical dense globular cluster) 
to be $f_{\rm b}\approx 0.8\%$ (as the fraction of the all objects in the core).
Adopting $a_{\rm pre} \approx  1 R_\odot$, we find 
that the formation rate of potential BH-WD X-ray binaries through exchanges 
coupled with TIMT is $\Gamma_{\rm exch,1} \approx 2 \times 10^{-3} $ per Gyr per BH.

Relatively wide hard BH-WD binaries ($a\sim 80-1000 R_\odot$) can be formed 
through encounters with both MS-WD and WD-WD binaries, 
as pre-encounter binaries can be up to 20 $R_\odot$ (here, MS-WD binaries could be only these that have $a\ga 15 R_\odot $). 
Again, using the same set of simulations, we find that such binaries have a relative fraction $f_{\rm b}=1.4\%$ and
an average  $a_{\rm pre} \approx  22 R_\odot$.
The resulting formation rate is fairly 
similar to the formation via encounters with WD-WD binaries, 
$\Gamma_{\rm exch,2} \approx 0.06 f_{\rm hard}\approx 2\times 10^{-3}$ per Gyr per BH.

In our ``optimistic'' case, when in all binaries with  $a_{\rm post} \la 80 R_\odot$ TIMT works,
the fraction of binaries playing role for direct exchanges (with separations $a\la 3.2 R_\odot=80 R_\odot/25$)
is $f_{\rm b}\sim 2\%$. When averaged over all binaries in several numerical simulations, $f_{\rm hard}\sim 25\%$ (for $f_{\rm BH,0.1}=1$).
We find then $\Gamma_{\rm exch,1}= 6\times 10^{-3} $ 
and $\Gamma_{\rm exch,2}= 1.5 \times 10^{-2}$ per Gyr per BH.

\subsection{Physical collisions}

Physical collisions between compact stars (neutron stars, NSs) and low-mass red-giants
can lead to a formation of NS-WD X-ray binaries \citep{ivanova05}, which, during their persistent phase, are also known as
ultracompact X-ray binaries (UCXBs).
From observations of UCXBs in the Milky Way's galactic globular clusters, we know that UCXBs are formed at a high rate.
The fraction of UCXBs among all LMXBs is much larger in globular clusters than in the field \citep{2000ApJ...530L..21D}.

When natal kicks from \cite{2005MNRAS.360..974H} are adopted
for NSs formed via core collapse and reduced kicks are adopted for NS formed via electron capture supernova, the theoretically predicted UCXB formation rate through physical collisions is consistent with the observations of globular clusters both in the Milky Way amd in external galaxies \citep{ivanova05}. 

Like in the case of UCXBs formation, it is plausible that BHs, if retained and not detached in a BHs subcluster from the rest of the stellar population,
will experience physical collisions with RGs. The rate of such collisions, per BH,  can be estimated as in \cite{ivanova05}:

\begin{eqnarray}
\Gamma_{\rm BHRG}&\approx& 2 \pi G f_{\rm RG}  f_{\rm p} m_{BH}  n_{\rm c}  {\bar R}_{\rm rg}  v_{\infty}^{-1}  \\
&\approx & 0.1  f_{\rm RG} f_{\rm p} 
\frac{m_{\rm BH}} {15 M_\odot} \frac{n_{\rm c}}{10^5 {\rm pc}^{-3}} \frac{\bar R_{\rm RG}}{R_\odot} \frac{10{\rm km/s}}{v_{\infty}} {\rm per \ Gyr} \nonumber
\end{eqnarray}
where ${\bar R_{\rm RG}}$ is the average radius of the red giants,
$f_{\rm p}= r_{\rm p}/{\bar R_{\rm RG}}$ describes how close has to be an encounter in order to result in the formation of a binary compact enough
to start the MT. For NSs, $f_{\rm p} \approx 1.3$ \citep{lombardi06}. 
However for BHs the maximum periastron that leads to RGs Roche lobe overflow at the closest approach is larger, $f_{\rm p} \approx 5$.\footnote {We will
show in the paper in prep, where BH-RG encounters are modeled using the SPH code described in \citet{lombardi06}, 
that this is indeed the case, and the collisions with $f_{\rm p}\la 5$ lead to the formation of bound binary systems.}
The fraction $f_{\rm RG}$ of stars  at the RG stage is typically $\sim 0.4\%$ of non-degenerate single stars at the age $10-12$ Gyr, 
with the minimum stellar mass in the population being $0.08M_\odot$ and with the initial mass function as in \citet{2002Sci...295...82K}.
Mass segregation of stars in the core results in a higher  $f_{\rm RG}$.
Analyzing the set of simulations from \cite{ivanova08}, we find that this fraction is about $0.8\%$ of the overall core population
and ${\bar R_{\rm RG}}=3.7$ at 10 Gyr; 
in the simulations presented in \cite{freg_bf}, the post-processed $f_{\rm RG}\approx 0.8\%$ as well.
In our relatively dense stellar cluster, with $n_{\rm c}=10^5$ pc$^{-3}$ and velocity dispersion of about 10 km/s, 
a BH of 15 $M_\odot$ therefore could have $\sim 3\times10^{-3} f_{\rm p}$ 
chance of a physical collisions during 1 Gyr. Therefore physical collisions can provide a significant contribution
to overall BH-WD X-ray binary formation only if all the collisions leading to the formation of bound binaries, 
with all the periastrons, up to $5 R_{\rm RG}$, lead to X-ray binary formation.

Let us estimate what fraction of the formed bound binaries can successfully become BH-WD X-ray binaries.
A simple estimate for the outcome of a BH-RG collision can be done using an energy argument, 
as in the standard common envelope consideration, using the $\alpha_{\rm CE}\lambda$ formalism \citep{webbink84}.  Specifically, the final semimajor axis $a_f$ is given by

\begin{equation}
a_{\rm f} =  {R_{\rm RG}} \frac{\alpha_{\rm CE} {\lambda}}{2} \frac{m_{\rm BH}}{m_{\rm RG}} \frac{ m_{\rm RG,core} }{ m_{\rm RG, env}} \ ,
\label{ace}
\end{equation}
where $m_{\rm RG,core}$ and  $m_{\rm RG, env}$ are respectively the core and envelope masses of the RG, and $R_{\rm RG}$ is the RG radius.
The paramter $\lambda$ introduced to characterize the donor envelope central concentration
can be found directly from stellar models and is about 1 for low-mass RGs. The parameter $\alpha_{\rm CE}$ is 
introduced as a measure of the energy transfer efficiency from the orbital energy 
into envelope expansion, and is bound to be below 1.
The energy balance in the eq.(\ref{ace}) assumes that the amount of the energy that is transferred from the orbital motion to the envelope expansion 
does not exceed the energy that is required to eject the envelope to infinity. The kinetic energy that the ejected gas has at infinity is neglected, as well as that two stars had initially at the infinity, as it is very small compared to the binding energy of the envelope.

It can be seen, if  $\alpha_{\rm CE}=1$, that the final binary separation for a case of a BH collision with a low-mass RG
will result in this case in a formation of a binary with $a_{\rm f}> 1.3 R_{\rm RG}$ (this minimum value is for a case of an early 
subgiant, when the core mass is minimum, e.g., it is $\sim 0.12 M_\odot$ for $0.9 M_\odot$ RG with $\sim 2 R_\odot$ radius and $z=0.005$). 
Comparison with the parameter space on Fig.\ref{tau} shows that, if a post-collision eccentricity is low,
only encounters with a subgiant $R_{\rm RG} { m_{\rm RG,core} }/{ m_{\rm RG, env}} \la 0.66 R_\odot$ could  lead directly to a BH-WD X-ray binary,
this corresponds to a subgiant with $m_{\rm RG,core}\approx 0.17 M_\odot$  (considering a $0.9 M_\odot$ RG; its radius then is only $2.75 R_\odot$).
The time this RG spends as a subgiant before it reaches  $m_{\rm RG,core}\approx 0.17 M_\odot$ is only 135 Myr, or $\sim 25\%$
of its RG lifetime. Assuming that the final separation after a physical collision will be not exceeding the predicted by
simple energy conservation, the formation rates are then $\sim  2 \times 10^{-3}f_{\rm p}$ per BH per Gyr.
This rate is an upper limit, as it also is being assumed that all the physical collisions that lead to the formation of bound systems will
lead as well to the RG envelope stripping. From SPH simulations, we find that collisions with $f_{\rm p}\ga1$ will lead 
to a formation of binaries with $e\la 0.8$, making $f_{\rm p}\approx 1$ to be a border line between these collisions that
lead to the formation of BH-WD binaries able to start MT in isolation, and these that 
could not (see also Fig.1).\footnote{Details 
of the hydrodynamical simulations of collisions between BH and RG 
will be presented in the paper in prep.} This non-TIMT channel therefore implies a formation rate of $\sim  4 \times 10^{-4}$ per BH per Gyr.

For collisions with $f_{\rm p}\ga 1$, TIMT is the main mechanism that brings a formed binary to the MT.
We can estimate the maximum optimistic fraction. For that, we find what fraction of collision would lead
to the formation of BH-WD binaries with $a\la 80 R_\odot$, assuming that, even if a physical collision itself could not form
a binary that will start MT in isolation, TIMT will help (accordingly, $f_{\rm p}=5$).
Using the eq.(\ref{ace}), we find that the RG should have $R_{\rm RG} { m_{\rm RG,core} }/{ m_{\rm RG, env}} \la 9.6 R_\odot$.
This occurs when a $0.9 M_\odot$ RG has  $m_{\rm RG,core}\approx 0.31 M_\odot$ and  $R_{\rm RG}\approx 17.8 R_\odot$
and corresponds to almost 95\% RG lifetime (accordingly, and roughly, all collisions with RG can be counted
towards TIMT influenced collisions).
We conclude therefore that most optimistic formation rate via physical collisions, when TIMT is taken into account, is $1.5\times 10^{-2}$ per BH per Gyr.

\subsection{Three-body binary formation}

Three-body binary formation occurs if 

\begin{itemize}
\item three objects will meet within some vicinity of each other $a_{\rm v}$;
\item in the case of such a meeting at least two of the objects will form a bound system;
\item the formation of this bound system is not altered by physical collisions or strong tidal interactions between the objects
due to finite size effects.
\end{itemize}

In the literature, usually only the first condition is meant when generalised 
three-body binary formation rate is described \citep[e.g., ][]{bt, bf_nata}.
There are however limitations that make such treatments inapplicable to our case.
The formation rate in \cite{bt} is derived for equal masses only and is derived to estimate 
only the formation of binaries which have the hardness ratio $\eta$ of their binding energy
to the kinetic energy of an average object in the core to be 1.
In \cite{bf_nata}, the formation rate is derived to take into account unequal masses and different energies;
however, the treatment is applicable only when the resulting binaries are near the hard-soft boundary or a bit harder.

In their derivation of the formation rates of binaries with different energies,
\cite{bf_nata},  
approximate that the orbital separation in the formed binary is the same as 
the size of the vicinity $a_{\rm v}$ where the three objects meet. 
As shown by numerical experiments in \cite{aar76}, this assumption is approximately satisfied at least for equal masses,
where $a(1+e)/a_{\rm v}\approx 1$. 

In our study, we are most interested in the formation rates where objects 
have a fairly large mass ratio and have high energies ($\eta\ga100$).
Compared to $\eta=1$ binaries, the decrease in the hard binary formation rate due to physical collisions and tidal effects could be significant.
We therefore limit ourselves to the consideration of only degenerate objects meeting each other.
On the other hand, the limitation of the considered cases to only very hard binaries eliminates the necessity to consider the second condition: 
\cite{aar76} showed that the probability to form a binary is strongly increasing as vicinity is decreasing, $P\propto \eta^{-2}$, 
and, even in the case of $\eta = 2$, 77\% of three-body encounters resulted in binary formation.
They have also shown that the average eccentricity of formed binaries is a bit higher than in the thermal distribution:
$<e>=0.77$ for  $\eta = 2$ cases, though this was the largest energy considered and the average values were slowly decreasing
with hardness. We may expect that the average eccentricities could go as low as that of the thermal distribution. 

The rate at which three objects meet in the same vicinity  $a_{\rm v}$
is the product of the rate $\Gamma_2 (a_{\rm v}, m_1, m_2)$ for two objects to meet in
this neighborhood 
 and the probability $P_3$ that during this event a third object will pass by.
As previously, the two-body encounter rate $\Gamma_2 (a_{\rm v}, m_1, m_2)$ 
is standardly derived as a combination of geometric cross-section and gravitational focusing:

\begin{equation}
\Gamma_2 (a_{\rm v}, m_1, m_2) = \pi n_2 v_{\infty} a_{\rm v}^2 (1 + \frac{v_{\rm p}^2}{v_{\infty}^2}) \ ,
\end{equation}
where $v_{\rm p}^2=2G(m_1+m_2)/a_{\rm v}$.
Two objects spend in this vicinity about $\tau_{\rm v}\approx 2 a_{\rm v} / v_{\rm p}$.

The probability that a third object will be within the same vicinity, $P_3$, is then usually found only assuming that a third object will 
geometrically sweep a certain volume during $\tau_{\rm v}$ \citep{bt, bf_nata}. 
A ``geometric'' cross-section however is good only for cases of $a_{\rm v}$ large enough so $v_{\rm p} \la v_{\infty}$.
In the case when we are interested in the encounters occurring within a small vicinity only, gravitational focusing must be taken into account.
In some sense, this implies that we are looking for a probability that, at the same time, two objects will be gravitationally focused by
a BH. The probability that a third object will pass within  $a_{\rm v}$ during  $\tau_{\rm v}$ is then

\begin{equation}
P_3  \simeq 2 \pi n_3 a_{\rm v}^3  \frac{v_{\rm p}}{v_{\infty}}.
\end{equation}
The three-body formation rate then is

\begin{equation}
\Gamma_3 (a_{\rm v}) \simeq 2 \pi^2 n_2 n_3 a_{\rm v}^5    \frac{v_{\rm p}^3}{v_{\infty}^2} \simeq
2^{5/2} \pi^2 G^{3/2} n_2 n_3 a_v^{3.5} v_{\infty}^{-2}  m_{\rm BH} ^{3/2}.
\end{equation}

For very hard binaries, this rate is significantly larger than the corresponding rate derived in \cite{bf_nata} and can be written as

\begin{eqnarray}
&\Gamma_3&(a_{\rm v}) \simeq 1.6\times 10^{-16} {\rm per\ Gyr} \times \\
 & &\left(\frac{n_{\rm c}}{10^5 {\rm pc}^{-3}}\right)^2 \left(\frac{a_v}{R_\odot}\right)^{3.5} \left(\frac{10 {\rm km/s}}{v_{\infty}}\right)^{2} 
\left( \frac{m_{\rm BH}}{15 M_\odot}\right)^{3/2}.  \nonumber 
\end{eqnarray}

Therefore, even in a relatively dense stellar cluster, 
a BH of 15 $M_\odot$ will have only about $2\times10^{-9}$ three-body encounters within its 100 $R_\odot$ vicinity during 1 Gyr.
The formation rate necessary to explain BH-WD X-ray binaries formation is consistent only with encounters within
$\sim 5000 R_\odot$, or $\eta=10$. 
We also note that in this case $v_{\rm p}\gg v_\infty $, the assumption used for the derivation, is no longer valid.
Even in a denser cluster ($n_{\rm c}\sim 10^6 {\rm pc}$) with a somewhat smaller velocity dispersion, the rate given by the above estimate is still about an order of magnitude
less than the observed formation rate.

We conclude that unless a significant fraction of small-vicinity three-body encounters lead to a formation of binaries much smaller than the vicinity
where they met  (10-50 times smaller), three-body binary formation can not explain the observed rates of BH-WD X-ray binary formation.

\section{Conclusions}

\begin{table}
\caption{Formation rates}
\begin{tabular}{|l | c | c | }
\hline 
Channel & Conservative & Optimistic \\
\hline 
EX + TIMT & $2\times 10^{-3}$ & $6\times 10^{-3}$\\
EX + HARD + TIMT &$2\times 10^{-3}$ & $1.5\times10^{-2}$\\
PC & $4\times 10^{-4}$ & --  \\
PC +TIMT & -- & $1.5\times 10^{-2}$\\
\hline
\end{tabular}

\label{table_rates}
\footnotesize{The table shows the formation rates of BH-WD MT binaries,  
per BH per Gyr. Here ``EX + TIMT'' is formation via an exchange encounter 
with the consequent TIMT (triple-induced mass transfer), ``EX + HARD + TIMT'' is an exchange with a wide binary formation with the consequent 
hardening and TIMT. ``PC'' is a physical collision directly leading to MT, and ``PC + TIMT'' is 
the formation through a physical collisions of a bound but wide BH-WD binary that is brought to MT by TIMT.
``Conservative'' is the case when we take into account only BH-WDs with $a\la 35 R_\odot$, which
are guaranteed to create Kozai triple that results in TIMT during 1 Gyr. 
``Optimistic'' is the case when all BH-WD binaries with  $a\la 80 R_\odot$
have a chance to be in such Kozai triple that leads to TIMT, a process that takes several Gyr.}
\end{table}

We have discussed which BH-WD binaries can become X-ray binaries, how these binaries can be formed and at what rate.
We have considered the following formation channels:
\begin{itemize}
\item exchange encounters 
\item physical collisions with giants
\item three body binary formation
\end{itemize}
All of the channels require that at least a fraction of BHs interacts strongly with other stars in the cluster.

We find that the most important mechanism to make a BH-WD X-ray binary from an initially dynamically formed BH-WD binary 
is the triple induced mass transfer (TIMT), 
the second most important is hardening with the consequent TIMT (see Table~2).
TIMT is found to be very effective for BH-WD binaries and is not expected to be effective at all for BH-MS binaries,
as most of potential triple formation encounters will result in BH-MS collision.
We do not find physical collisions, that form MT binaries directly, with no other strong encounter, to play a significant role.

With regard to TIMT, we distinguish between ``conservative'' and ``optimstic'' scenario.
In the conservative case, we take into account only BH-WD binaries with  $a\la 35 R_\odot$. These binaries
are essentially guaranteed to undergo TIMT within 1 Gyr. In the optimitic case, we take into account
all BH-WD binaries with  $a\la 80 R_\odot$. For these binaries, TIMT has a time-scale of several Gyr.

With our conservative estimate of the formation rates, we find that we can explain the formation rate inferred from observations, $\sim 4\times 10^{-3}/f_{\rm BH,0.1}$,
but only if $f_{\rm BH,0.1}\approx 1$.
This, by our definition of $f_{\rm BH,0.1}$, corresponds the case when 10\% of all formed BHs remain in the cluster
and interact with the core's stellar populations.

With the optimistic scenario, we find that all the channels have comparable formation rates and
each of the channels can explain the observed formation rate of BH-WD X-ray binaries.
The combined rate of all the channels is $\sim 4\times 10^{-2}$ per BH per Gyr.
Comparing this with the formation rates inferred from the observations,
we find that we can explain the observations only with $f_{\rm BH}\ga 0.1$.
In other words, even in the most optimistic case, we require that at least
1\% of all initially formed BHs ($f_{\rm BH,0.1}=0.1$) should be both not evaporated from the cluser  and not 
dynamically detached from the core's stellar population to BH subcluster.
Future simulations could address this.

\acknowledgments

JMF acknowledges support from 
Chandra/Einstein Postdoctoral Fellowship Award PF7-80047.
NI, SC, COH, \& TW acknowledge support from NSERC.

\bibliography{bhwd}

\begin{thebibliography}{53}
\expandafter\ifx\csname natexlab\endcsname\relax\def\natexlab#1{#1}\fi

\bibitem[{{Aarseth} \& {Heggie}(1976)}]{aar76}
{Aarseth}, S.~J., \& {Heggie}, D.~C. 1976, \aap, 53, 259

\bibitem[{{Albrow} {et~al.}(2001){Albrow}, {Gilliland}, {Brown}, {Edmonds},
  {Guhathakurta}, \& {Sarajedini}}]{Albrow01}
{Albrow}, M.~D., {Gilliland}, R.~L., {Brown}, T.~M., {Edmonds}, P.~D.,
  {Guhathakurta}, P., \& {Sarajedini}, A. 2001, \apj, 559, 1060

\bibitem[{{Angelini} {et~al.}(2001){Angelini}, {Loewenstein}, \&
  {Mushotzky}}]{2001ApJ...557L..35A}
{Angelini}, L., {Loewenstein}, M., \& {Mushotzky}, R.~F. 2001, \apjl, 557, L35

\bibitem[{{Belczynski} {et~al.}(2006){Belczynski}, {Sadowski}, {Rasio}, \&
  {Bulik}}]{2006ApJ...650..303B}
{Belczynski}, K., {Sadowski}, A., {Rasio}, F.~A., \& {Bulik}, T. 2006, \apj,
  650, 303

\bibitem[{{Binney} \& {Tremaine}(1987)}]{bt}
{Binney}, J., \& {Tremaine}, S. 1987, {Galactic dynamics}, ed. S.~Binney, J.
  \&~Tremaine

\bibitem[{{Deutsch} {et~al.}(2000){Deutsch}, {Margon}, \&
  {Anderson}}]{2000ApJ...530L..21D}
{Deutsch}, E.~W., {Margon}, B., \& {Anderson}, S.~F. 2000, \apjl, 530, L21

\bibitem[{{Di Stefano} {et~al.}(2002){Di Stefano}, {Kong}, {Garcia}, {Barmby},
  {Greiner}, {Murray}, \& {Primini}}]{2002ApJ...570..618D}
{Di Stefano}, R., {Kong}, A.~K.~H., {Garcia}, M.~R., {Barmby}, P., {Greiner},
  J., {Murray}, S.~S., \& {Primini}, F.~A. 2002, \apj, 570, 618

\bibitem[{{Downing} {et~al.}(2009){Downing}, {Benacquista}, {Giersz}, \&
  {Spurzem}}]{2009arXiv0910.0546D}
{Downing}, J.~M.~B., {Benacquista}, M.~J., {Giersz}, M., \& {Spurzem}, R. 2009,
  ArXiv e-prints

\bibitem[{{Eggleton} \& {Kiseleva-Eggleton}(2001)}]{Eggleton01_triples}
{Eggleton}, P.~P., \& {Kiseleva-Eggleton}, L. 2001, \apj, 562, 1012

\bibitem[{{Fregeau} {et~al.}(2004){Fregeau}, {Cheung}, {Portegies Zwart}, \&
  {Rasio}}]{freg04}
{Fregeau}, J.~M., {Cheung}, P., {Portegies Zwart}, S.~F., \& {Rasio}, F.~A.
  2004, \mnras, 352, 1

\bibitem[{{Fregeau} {et~al.}(2009{\natexlab{a}}){Fregeau}, {Ivanova}, \&
  {Rasio}}]{freg_bf}
{Fregeau}, J.~M., {Ivanova}, N., \& {Rasio}, F.~A. 2009{\natexlab{a}}, ArXiv
  e-prints

\bibitem[{{Fregeau} {et~al.}(2009{\natexlab{b}}){Fregeau}, {Richer}, {Rasio},
  \& {Hurley}}]{2009ApJ...695L..20F}
{Fregeau}, J.~M., {Richer}, H.~B., {Rasio}, F.~A., \& {Hurley}, J.~R.
  2009{\natexlab{b}}, \apjl, 695, L20

\bibitem[{{Gehrels}(1986)}]{1986ApJ...303..336G}
{Gehrels}, N. 1986, \apj, 303, 336

\bibitem[{{Gnedin} {et~al.}(2009){Gnedin}, {Maccarone}, {Psaltis}, \&
  {Zepf}}]{2009ApJ...705L.168G}
{Gnedin}, O.~Y., {Maccarone}, T.~J., {Psaltis}, D., \& {Zepf}, S.~E. 2009,
  \apjl, 705, L168

\bibitem[{{Heggie} \& {Hut}(2003)}]{heggiehut_03}
{Heggie}, D., \& {Hut}, P. 2003, {The Gravitational Million-Body Problem: A
  Multidisciplinary Approach to Star Cluster Dynamics}, ed. P.~Heggie, D.
  \&~Hut

\bibitem[{{Heggie}(1975)}]{heggie75}
{Heggie}, D.~C. 1975, \mnras, 173, 729

\bibitem[{{Heggie} {et~al.}(1996){Heggie}, {Hut}, \&
  {McMillan}}]{1996ApJ...467..359H}
{Heggie}, D.~C., {Hut}, P., \& {McMillan}, S.~L.~W. 1996, \apj, 467, 359

\bibitem[{{Hobbs} {et~al.}(2005){Hobbs}, {Lorimer}, {Lyne}, \&
  {Kramer}}]{2005MNRAS.360..974H}
{Hobbs}, G., {Lorimer}, D.~R., {Lyne}, A.~G., \& {Kramer}, M. 2005, \mnras,
  360, 974

\bibitem[{{Humphrey} \& {Buote}(2008)}]{2008ApJ...689..983H}
{Humphrey}, P.~J., \& {Buote}, D.~A. 2008, \apj, 689, 983

\bibitem[{{Hut}(1983)}]{hut83}
{Hut}, P. 1983, \apjl, 272, L29

\bibitem[{{Hut} \& {Bahcall}(1983)}]{hutbah83}
{Hut}, P., \& {Bahcall}, J.~N. 1983, \apj, 268, 319

\bibitem[{{Innanen} {et~al.}(1997){Innanen}, {Zheng}, {Mikkola}, \&
  {Valtonen}}]{Innanen97_kozai}
{Innanen}, K.~A., {Zheng}, J.~Q., {Mikkola}, S., \& {Valtonen}, M.~J. 1997,
  \aj, 113, 1915

\bibitem[{{Irwin} {et~al.}(2009){Irwin}, {Brink}, {Bregman}, \&
  {Roberts}}]{2009arXiv0908.1115I}
{Irwin}, J.~A., {Brink}, T., {Bregman}, J.~N., \& {Roberts}, T.~P. 2009, ArXiv
  e-prints

\bibitem[{{Ivanova}(2008{\natexlab{a}})}]{natatrip08}
{Ivanova}, N. 2008{\natexlab{a}}, in Multiple Stars Across the H-R Diagram, ed.
  {S.~Hubrig, M.~Petr-Gotzens, \& A.~Tokovinin}, 101--+

\bibitem[{{Ivanova}(2008{\natexlab{b}})}]{natatrip}
{Ivanova}, N. 2008{\natexlab{b}}, in Multiple Stars Across the H-R Diagram, ed.
  {S.~Hubrig, M.~Petr-Gotzens, \& A.~Tokovinin}, 101--+

\bibitem[{{Ivanova} {et~al.}(2005{\natexlab{a}}){Ivanova}, {Belczynski},
  {Fregeau}, \& {Rasio}}]{bf_nata}
{Ivanova}, N., {Belczynski}, K., {Fregeau}, J.~M., \& {Rasio}, F.~A.
  2005{\natexlab{a}}, \mnras, 358, 572

\bibitem[{{Ivanova} {et~al.}(2008){Ivanova}, {Heinke}, {Rasio}, {Belczynski},
  \& {Fregeau}}]{ivanova08}
{Ivanova}, N., {Heinke}, C.~O., {Rasio}, F.~A., {Belczynski}, K., \& {Fregeau},
  J.~M. 2008, \mnras, 386, 553

\bibitem[{{Ivanova} {et~al.}(2006){Ivanova}, {Heinke}, {Rasio}, {Taam},
  {Belczynski}, \& {Fregeau}}]{nata_wd_06}
{Ivanova}, N., {Heinke}, C.~O., {Rasio}, F.~A., {Taam}, R.~E., {Belczynski},
  K., \& {Fregeau}, J. 2006, \mnras, 372, 1043

\bibitem[{{Ivanova} {et~al.}(2005{\natexlab{b}}){Ivanova}, {Rasio}, {Lombardi},
  {Dooley}, \& {Proulx}}]{ivanova05}
{Ivanova}, N., {Rasio}, F.~A., {Lombardi}, Jr., J.~C., {Dooley}, K.~L., \&
  {Proulx}, Z.~F. 2005{\natexlab{b}}, \apjl, 621, L109

\bibitem[{{Kalogera} {et~al.}(2004){Kalogera}, {King}, \&
  {Rasio}}]{kalogera_04}
{Kalogera}, V., {King}, A.~R., \& {Rasio}, F.~A. 2004, \apjl, 601, L171

\bibitem[{{Kim} {et~al.}(2006){Kim}, {Kim}, {Fabbiano}, {Lee}, {Park},
  {Geisler}, \& {Dirsch}}]{kim06}
{Kim}, E., {Kim}, D., {Fabbiano}, G., {Lee}, M.~G., {Park}, H.~S., {Geisler},
  D., \& {Dirsch}, B. 2006, \apj, 647, 276

\bibitem[{{Kozai}(1962)}]{kozai_62}
{Kozai}, Y. 1962, \aj, 67, 591

\bibitem[{{Kroupa}(2002)}]{2002Sci...295...82K}
{Kroupa}, P. 2002, Science, 295, 82

\bibitem[{{Kulkarni} {et~al.}(1993){Kulkarni}, {Hut}, \&
  {McMillan}}]{1993Natur.364..421K}
{Kulkarni}, S.~R., {Hut}, P., \& {McMillan}, S. 1993, Nature, 364, 421

\bibitem[{{Kundu} {et~al.}(2002){Kundu}, {Maccarone}, \&
  {Zepf}}]{2002ApJ...574L...5K}
{Kundu}, A., {Maccarone}, T.~J., \& {Zepf}, S.~E. 2002, \apjl, 574, L5

\bibitem[{{Lombardi} {et~al.}(2006){Lombardi}, {Proulx}, {Dooley}, {Theriault},
  {Ivanova}, \& {Rasio}}]{lombardi06}
{Lombardi}, Jr., J.~C., {Proulx}, Z.~F., {Dooley}, K.~L., {Theriault}, E.~M.,
  {Ivanova}, N., \& {Rasio}, F.~A. 2006, \apj, 640, 441

\bibitem[{{Maccarone} {et~al.}(2007){Maccarone}, {Kundu}, {Zepf}, \&
  {Rhode}}]{2007Natur.445..183M}
{Maccarone}, T.~J., {Kundu}, A., {Zepf}, S.~E., \& {Rhode}, K.~L. 2007, Nature,
  445, 183

\bibitem[{{Mikkola}(1983)}]{mikkola83}
{Mikkola}, S. 1983, \mnras, 203, 1107

\bibitem[{{Miller} \& {Hamilton}(2002{\natexlab{a}})}]{Miller02_kozai}
{Miller}, M.~C., \& {Hamilton}, D.~P. 2002{\natexlab{a}}, \apj, 576, 894

\bibitem[{{Miller} \& {Hamilton}(2002{\natexlab{b}})}]{2002MNRAS.330..232C}
---. 2002{\natexlab{b}}, \mnras, 330, 232

\bibitem[{{Milone} {et~al.}(2008){Milone}, {Piotto}, {Bedin}, \&
  {Sarajedini}}]{Milone08}
{Milone}, A.~P., {Piotto}, G., {Bedin}, L.~R., \& {Sarajedini}, A. 2008,
  Memorie della Societa Astronomica Italiana, 79, 623

\bibitem[{{O'Leary} {et~al.}(2006){O'Leary}, {Rasio}, {Fregeau}, {Ivanova}, \&
  {O'Shaughnessy}}]{oleary_06}
{O'Leary}, R.~M., {Rasio}, F.~A., {Fregeau}, J.~M., {Ivanova}, N., \&
  {O'Shaughnessy}, R. 2006, \apj, 637, 937

\bibitem[{{Peters}(1964)}]{peters64}
{Peters}, P.~C. 1964, Physical Review, 136, 1224

\bibitem[{{Portegies Zwart} \& {McMillan}(2000)}]{2000ApJ...528L..17P}
{Portegies Zwart}, S.~F., \& {McMillan}, S.~L.~W. 2000, \apjl, 528, L17

\bibitem[{{Rosswog} {et~al.}(2009){Rosswog}, {Ramirez-Ruiz}, \&
  {Hix}}]{2009ApJ...695..404R}
{Rosswog}, S., {Ramirez-Ruiz}, E., \& {Hix}, W.~R. 2009, \apj, 695, 404

\bibitem[{{Roy} \& {Haddow}(2003)}]{2003CeMDA..87..411R}
{Roy}, A., \& {Haddow}, M. 2003, Celestial Mechanics and Dynamical Astronomy,
  87, 411

\bibitem[{{Sarazin} {et~al.}(2003){Sarazin}, {Kundu}, {Irwin}, {Sivakoff},
  {Blanton}, \& {Randall}}]{2003ApJ...595..743S}
{Sarazin}, C.~L., {Kundu}, A., {Irwin}, J.~A., {Sivakoff}, G.~R., {Blanton},
  E.~L., \& {Randall}, S.~W. 2003, \apj, 595, 743

\bibitem[{{Sigurdsson} \& {Hernquist}(1993)}]{1993Natur.364..423S}
{Sigurdsson}, S., \& {Hernquist}, L. 1993, Nature, 364, 423

\bibitem[{{Spitzer}(1969)}]{1969ApJ...158L.139S}
{Spitzer}, L.~J. 1969, \apjl, 158, L139+

\bibitem[{{Verbunt} \& {Lewin}(2006)}]{2006csxs.book..341V}
{Verbunt}, F., \& {Lewin}, W.~H.~G. 2006, {Globular cluster X-ray sources}, ed.
  M.~Lewin, W. H. G. \& van der~Klis, 341--379

\bibitem[{{Watters} {et~al.}(2000){Watters}, {Joshi}, \&
  {Rasio}}]{2000ApJ...539..331W}
{Watters}, W.~A., {Joshi}, K.~J., \& {Rasio}, F.~A. 2000, \apj, 539, 331

\bibitem[{{Webbink}(1984)}]{webbink84}
{Webbink}, R.~F. 1984, \apj, 277, 355

\bibitem[{{Zepf} {et~al.}(2008){Zepf}, {Stern}, {Maccarone}, {Kundu},
  {Kamionkowski}, {Rhode}, {Salzer}, {Ciardullo}, \&
  {Gronwall}}]{2008ApJ...683L.139Z}
{Zepf}, S.~E., {Stern}, D., {Maccarone}, T.~J., {Kundu}, A., {Kamionkowski},
  M., {Rhode}, K.~L., {Salzer}, J.~J., {Ciardullo}, R., \& {Gronwall}, C. 2008,
  \apjl, 683, L139

\end{thebibliography}
\bibliographystyle{apj}

\end{document}